\documentclass[twocolumn, tighten]{aastex63}

\usepackage{nicefrac}
\usepackage{microtype}
\usepackage{booktabs}
\usepackage{rotating}
\usepackage{xcolor}

\newcommand{\lya}{Ly$\alpha$}

\newcommand{\kms}{${\rm km\,s}^{-1}$}

\newcommand{\es}{${\rm 1ES}\,1553{+}113$}

\received{XX XX, 2019}
\revised{XX XX, 2019}
\accepted{XX}
\submitjournal{ApJL}

\begin{document}

\title{The Physical Origins of the Identified and Still Missing Components of the Warm-Hot Intergalactic Medium: Insights from Deep Surveys in the Field of Blazar 1ES1553+113}

\correspondingauthor{Sean D. Johnson}
\email{sdj@astro.princeton.edu}

\author[0000-0001-9487-8583]{Sean D. Johnson}
\altaffiliation{Hubble \& Carnegie-Princeton fellow}
\affil{Department of Astrophysical Sciences, 4 Ivy Lane, Princeton University, Princeton, NJ 08544, USA}
\affil{The Observatories of the Carnegie Institution for Science, 813 Santa Barbara Street, Pasadena, CA 91101, USA}

\author[0000-0003-2083-5569]{John S. Mulchaey}
\affil{The Observatories of the Carnegie Institution for Science, 813 Santa Barbara Street, Pasadena, CA 91101, USA}

\author[0000-0001-8813-4182]{Hsiao-Wen Chen}
\affil{Department of Astronomy \& Astrophysics, The University of Chicago, 5640 S. Ellis Avenue, Chicago, IL 60637, USA}

\author[0000-0001-6374-7185]{Nastasha A. Wijers}
\affil{Leiden Observatory, Leiden University, PO Box 9513, NL-2300 RA Leiden, the Netherlands}

\author[0000-0002-7898-7664]{Thomas Connor}
\affil{The Observatories of the Carnegie Institution for Science, 813 Santa Barbara Street, Pasadena, CA 91101, USA}

\author[0000-0001-5020-9994]{Sowgat Muzahid}
\affil{Leiden Observatory, Leiden University, PO Box 9513, NL-2300 RA Leiden, the Netherlands}

\author[0000-0002-0668-5560]{Joop Schaye}
\affil{Leiden Observatory, Leiden University, PO Box 9513, NL-2300 RA Leiden, the Netherlands}

\author{Renyue Cen}
\affil{Department of Astrophysical Sciences, 4 Ivy Lane, Princeton University, Princeton, NJ 08544, USA}

\author[0000-0002-5382-2898]{Scott G. Carlsten}
\affil{Department of Astrophysical Sciences, 4 Ivy Lane, Princeton University, Princeton, NJ 08544, USA}

\author{Jane Charlton}
\affil{Dept. of Astronomy \& Astrophysics, The Pennsylvania State University, 525 Davey Lab, University Park, PA 16802, USA}

\author[0000-0001-7081-0082]{Maria R. Drout}
\affil{Department of Astronomy and Astrophysics, University of Toronto, 50 St. George Street, Toronto, Ontario, M5S, 3H4 Canada}
\affil{The Observatories of the Carnegie Institution for Science, 813 Santa Barbara Street, Pasadena, CA 91101, USA}

\author{Andy D. Goulding}
\affil{Department of Astrophysical Sciences, 4 Ivy Lane, Princeton University, Princeton, NJ 08544, USA}

\author[0000-0001-6154-8983]{Terese T. Hansen}
\affil{Mitchell Institute for Fundamental Physics and Astronomy and Department of Physics and Astronomy, Texas A\&M University, College Station, TX~77843-4242, USA}

\author[0000-0002-6313-6808]{Gregory L. Walth}
\affil{The Observatories of the Carnegie Institution for Science, 813 Santa Barbara Street, Pasadena, CA 91101, USA}


\begin{abstract}

The relationship between galaxies and the state/chemical enrichment of the warm-hot intergalactic medium (WHIM) expected to dominate the baryon budget at low-$z$ provides sensitive constraints on structure formation and galaxy evolution models. We present a deep redshift survey in the field of 1ES1553$+$113, a blazar with a unique combination of UV$+$X-ray spectra for surveys of the circum-/intergalactic medium (CGM/IGM).  \cite{Nicastro:2018} reported the detection of two O\,VII WHIM absorbers at $z=0.4339$ and $0.3551$ in its spectrum, suggesting that the WHIM is metal-rich and sufficient to close the missing baryons problem. Our survey indicates that the blazar is a member of a $z=0.433$ group and that the higher-$z$ O\,VII candidate arises from its intragroup medium. The resulting bias precludes its use in baryon censuses. The $z=0.3551$ candidate occurs in an isolated environment $630$ kpc from the nearest galaxy (with stellar mass $\log M_*/M_\odot \approx 9.7$) which we show is unexpected for the WHIM.  Finally, we characterize the galactic environments of broad H\,I Ly$\alpha$ absorbers (Doppler widths of $b=40-80$ \kms; $T\lesssim4\times10^5$ K) which provide metallicity independent WHIM probes. On average, broad \lya\, absorbers are ${\approx}2\times$ closer to the nearest luminous ($L>0.25 L_*$) galaxy (700 kpc) than narrow ($b<30$ \kms; $T\lesssim4\times10^5$ K) ones (1300 kpc) but ${\approx}2\times$ further than O\,VI absorbers (350 kpc).  These observations suggest that gravitational collapse heats portions of the IGM to form the WHIM but with feedback that does not enrich the IGM far beyond galaxy/group halos to levels currently observable in UV/X-ray metal lines.

\end{abstract}

\keywords{intergalactic medium -- quasars: absorption lines -- BL Lacertae objects: 1ES 1553+113}

\section{Introduction} \label{section:intro}

Cosmological simulations predict that gravitational shocks associated with structure formation will heat a large fraction of the cool ($T\approx 10^4$ K) intergalactic medium (IGM) that dominates the baryon budget in the early Universe to form a Warm-Hot Intergalactic Medium (WHIM; $T\approx 10^5 - 10^7$ K) at  $z\lesssim1$ \citep[e.g.][]{Cen:1999}.  The predicted physical state and enrichment levels of the WHIM depend sensitively on stellar and black hole feedback which provide additional heating and chemical enrichment  \citep[e.g.][]{Rahmati:2016, Nelson:2018, Wijers:2019}. Observations of the WHIM and its relationship to galaxies can, therefore, serve as a check of our cosmological paradigm and as a laboratory for studying galaxy evolution.

While observationally elusive, the WHIM can be detected via absorption spectroscopy through ionic transitions {\color{black} in the UV and X-ray} as well as through metallicity independent probes such as broad H\,I \lya\ absorption \citep[e.g.][]{Danforth:2010}, the Sunyaev-Zel'dovich (SZ) effect \citep[e.g.][]{de-Graaff:2019}, and the dispersion measure of fast radio bursts \citep[FRBs; e.g.][]{Bannister:2019, Ravi:2019}. {\color{black} Surveys of the highly ionized phases of the CGM/IGM traced by O\,VI \citep[e.g.][]{Danforth:2016}, Ne\,VIII \citep[e.g.][]{Pachat:2017, Frank:2018}, and Mg\,X \citep[][]{Qu:2016} }with the Cosmic Origins Spectrograph \citep{Green:2012} on the {\it Hubble Space Telescope} (HST) can account for a large fraction of the baryons expected in the WHIM but leave $\sim30\%$ missing \citep[e.g.][]{Shull:2012a} and potentially in a chemically pristine or more highly ionized phase.

Surveys of CGM/IGM {\color{black} around galaxies} find that metal ion absorption is common in the CGM at projected distances ($d$) less than the estimated galaxy host halo virial radii ($R_{\rm h}$) but comparatively rare at larger distances \citep[e.g.][]{Liang:2014, Turner:2014, Johnson:2015a, Johnson:2017, Burchett:2019}. These observations suggest that feedback may be ineffective at enriching the IGM far beyond galaxy halos. Indeed, the statistical detection of SZ signal from the filaments between massive galaxies \citep[][]{de-Graaff:2019} can potentially account for the remaining missing baryons, suggesting that a substantial portion of the WHIM exhibits low metallicities \citep[$<\frac{1}{10}$ solar;][]{Liang:2014, Johnson:2015a} or high temperatures ($T>6\times10^{5}$ K) not traced in the UV.

New insights into chemical enrichment mechanisms and the physical state of the CGM/IGM require deep galaxy surveys in fields with UV and X-ray absorption spectra. Blazars are ideal for such studies because of their high UV/X-ray flux levels. Recently, \cite{Nicastro:2018} obtained a 1.7 Msec {\it XMM-Newton} X-ray spectrum of the blazar \es, reaching the S/N levels required to detect hot CGM/IGM absorbers individually over a large redshift pathlength for the first time. {\color{black} The X-ray spectrum revealed two candidate O\,VII absorption systems at $z=0.4339$ and $0.3551$, each with statistical significance of $\approx3-4\sigma$, though we note that systematic/non-Gaussian errors \citep[e.g.][]{Nevalainen:2019} and contamination \citep[e.g.][]{Nicastro:2016} have led to past controversies over X-ray absorbers. Nevertheless, taken together, the two O\,VII absorbers reported by \cite{Nicastro:2018} suggest that the hot phase of the CGM/IGM is metal-rich and accounts for $10-70\%$ of the  baryon budget.} However, the combination of a poorly constrained blazar redshift (due to a featureless spectrum) and limited complementary galaxy surveys complicates the interpretation of {\color{black} absorbers} toward \es.

Here, we present a deep and highly complete galaxy redshift survey in the field of \es. When combined with UV absorption spectra, the survey enables a precise measurement of the redshift of \es\ and provides insights into the origins of intervening IGM/CGM systems.  The letter proceeds as follows: In Section \ref{section:data}, we describe the galaxy survey and UV {\color{black} spectroscopy}. In Section \ref{section:redshift}, we combine these datasets to infer the blazar redshift. In Section \ref{section:environment}, we characterize the galactic environments of the candidate WHIM absorbers and draw insights into their origins.

{\color{black} We} adopt a flat $\Lambda$ cosmology with $\Omega_{\rm m}=0.3$, $\Omega_\Lambda = 0.7$, and $H_0=70\ {\rm km\,s^{-1}\,Mpc^{-1}}$. All  magnitudes are in the AB system. {\color{black} We define the knee in the galaxy luminosity function, $L_*$, as $M_r=-21.5$ \citep{Loveday:2012}}.

\section{Observations and Data}
\label{section:data}

\begin{figure*}
\includegraphics[width=\textwidth]{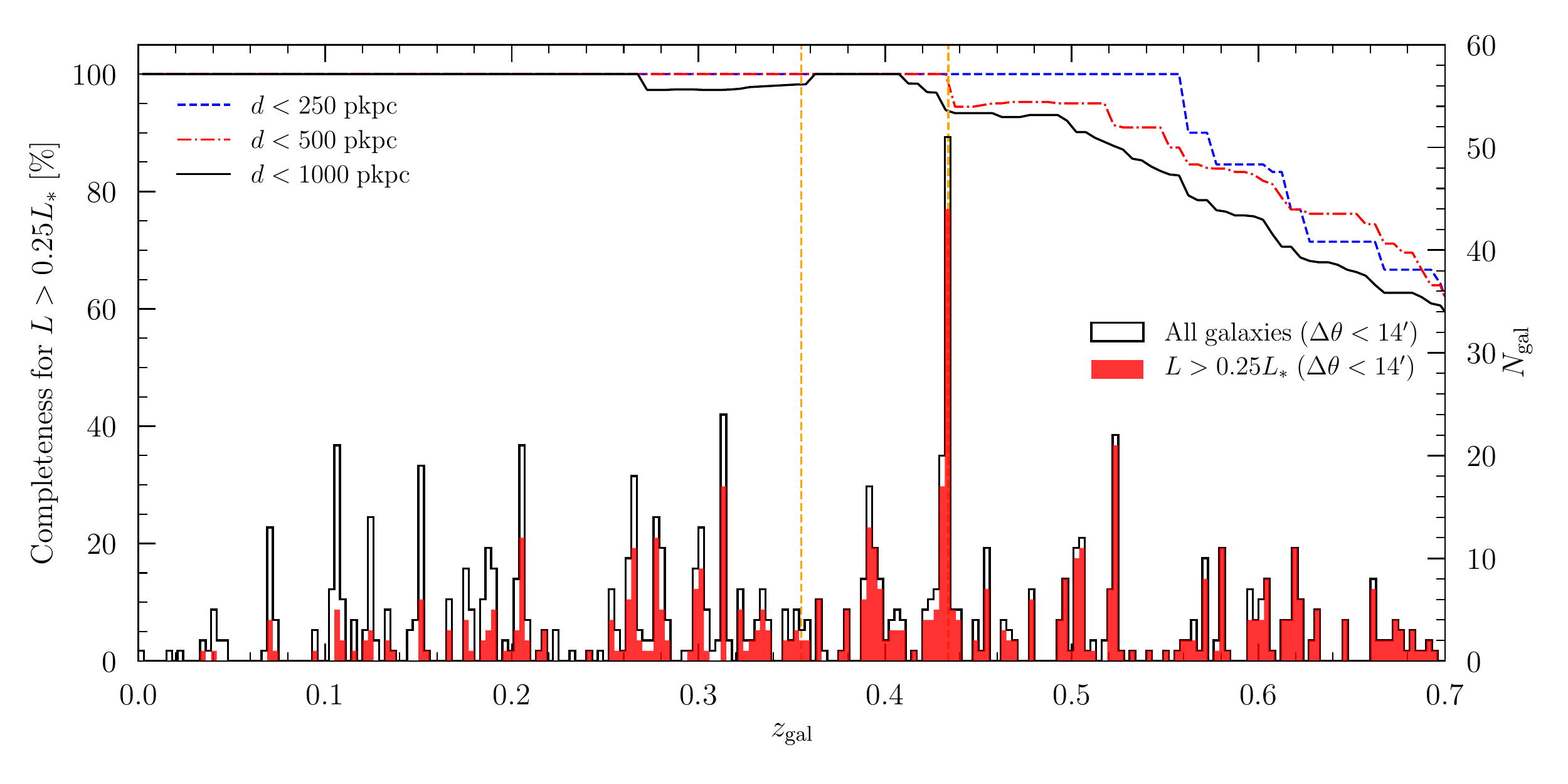}
\caption{{\it Left axis}: {\color{black} Redshift completeness for galaxies of $L>0.25 L_*$ versus redshift at projected distances of $d<250, 500,$ and $1000$ pkpc (top curves). {\it Right axis}: Redshift histograms with the full survey ($\Delta \theta < 14'$ from the blazar) in black and galaxies of $L>0.25 L_*$ in red. The redshifts of O\,VII candidates at $z=0.3551$ and $0.4339$ are shown in orange dashed lines.}}
\label{figure:completeness}
\end{figure*}

\subsection{Galaxy survey data}

To {\color{black} study} the relationship between galaxies and the IGM, we conducted a deep and highly complete redshift survey targeting galaxies of $m_r<23.5$ mag in the field of \es\ with {\color{black} multi-slit spectrographs} on the Magellan Telescopes. We acquired deep $g$-, $r$-, and $i$-band images with {\color{black} MOSAIC on the Mayall} telescope with 1800 sec of exposure in each filter under $1''$ {\color{black} seeing} (PI: Johnson; PID: 2015A-0187) and an {\it HST} image with the ACS$+$F814W filter and 1200 sec of exposure (PI: Mulchaey; PID: 13024). We processed the data as described in \cite{Chen:2009} and \cite{Johnson:2015a}. In total, we measured spectroscopic redshifts for 921 galaxies at angular distances of $\Delta \theta < 14'$ from the blazar sightline and also include 25 redshifts from \cite{Prochaska:2011} and one from \cite{Keeney:2018}. {\color{black} Redshift} histograms and completeness levels for galaxies of $L \gtrsim 0.25 L_*$ ($>90\%$ at projected distances of $d<500$ kpc and $z<0.5$) are shown in Figure \ref{figure:completeness}.

\begin{figure*}
\centering
    \includegraphics[width=\textwidth]{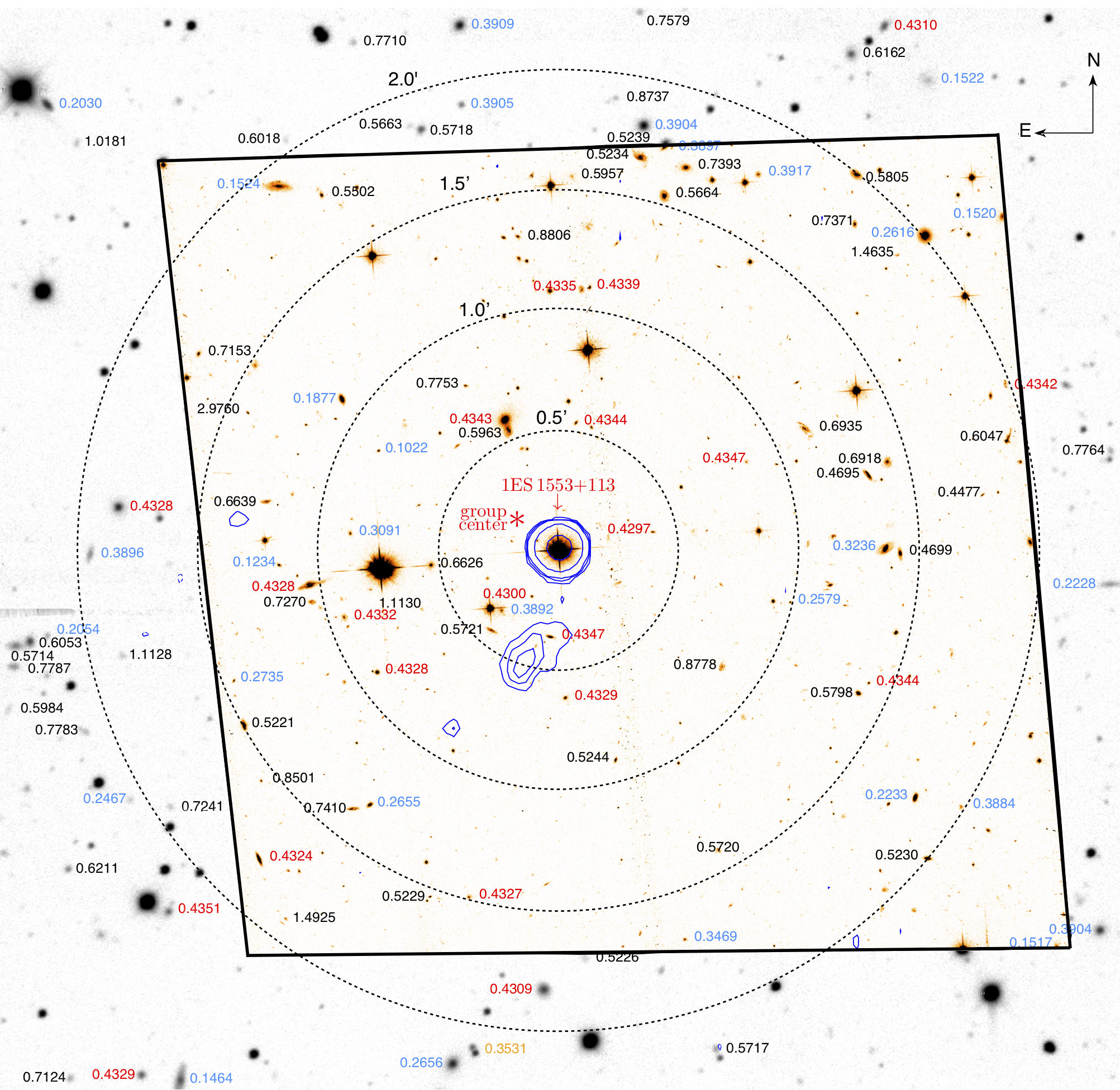}
    \caption{Image of the field of \es\ with galaxy redshifts labeled. The {\it HST} ACS$+$F814W image is shown in heat map while the outer regions not covered by the ACS are filled in with the MOSAIC $i$-band image shown in grey-scale. The galaxy labels are colored by redshift in black  ($z>0.4387$), red ($z=0.4291-0.4387$), orange ($z=0.3506-0.3596$), and blue (all other galaxies). The $z=0.4291-0.4387$ and  $z=0.3506-0.3596$ redshift intervals correspond to $\pm 1000$ \kms\ velocity intervals around the candidate O\,VII {\color{black} absorber redshifts}. The blue contours from the FIRST \citep[][]{Becker:1995} survey reveal a radio lobe. Dotted circles with radii of $0.5'$, $1.0'$, $1.5'$, and $2.0'$ are shown for scale (170, 340, 510, 680 pkpc at $z=0.433$).}
    \label{figure:field_image}
\end{figure*}

The {\color{black} survey} results are summarized in Table \ref{table:galaxies} which reports coordinates, apparent magnitudes ($m_g, m_r, m_i$), redshift quality (``g'' for secure redshifts and ``s'' for single-line redshifts), redshift ($z_{\rm gal}$), rest-frame $M_g - M_r$ color, absolute rest-frame $r$-band magnitude ($M_r$), stellar mass ($\log M_*/M_\odot$), and projected angular \& physical separations from the blazar sightline ($\Delta \theta$ \& $d$). The absolute magnitudes include $k$-corrections, and the stellar masses are {\color{black} estimated as in} \cite{Johnson:2015a} assuming a \cite{Chabrier:2003} IMF. Typical uncertainties in the redshifts, magnitudes, and stellar masses are $60$ \kms, $0.1$ mags, and $0.2$ dex respectively. Table \ref{table:galaxies} is separated into sections by redshift within $\pm1000$ \kms\ of the two candidate O\,VII absorbers ($z=0.4291-0.4387$; $0.3506-0.3596$), those at higher redshift ($z>0.4387$), and all other redshifts. Figure \ref{figure:field_image} displays an image of the field with galaxy redshifts labeled.

\subsection{UV absorption spectroscopy}

The COS GTO team acquired {\color{black} G130M$+$G160M} spectra of \es\ (PI: Green; PID: 11520, 12025) which are useful for inferring the redshift of the blazar based on the presence/absence of \lya\ forest absorption. We retrieved all available {\color{black} COS spectra} for \es\ from the {\it HST} archive and combined them into a single coadded spectrum as described in \cite{Johnson:2013}.

\begin{table*}
\caption{Redshift survey summary with galaxies separated by redshift. The full table is available on the journal webpage.}
\label{table:galaxies}
\centering

\begin{tabular}{ccccccccrrrr}

\hline 
R.A.    & Decl.   & $m_g$ & $m_r$ & $m_i$ & quality &     $z_{\rm gal}$ & $M_g - M_r$ & $M_r$ &   $\log M_*/M_\odot$           & \multicolumn{1}{c}{$\Delta \theta$} & \multicolumn{1}{c}{$d$} \\
(J2000) & (J2000) & (AB)  & (AB)  & (AB) &  & (AB)        & (AB)  &  &  & \multicolumn{1}{c}{(arcsec)}        & \multicolumn{1}{c}{(pkpc)} \\\hline \hline
\multicolumn{11}{c}{$z_{\rm gal}>0.4387$} \\
\hline
15:55:44.20 & +11:11:04.6 & 23.1 & 22.1 & 21.4 & g & 0.5721 & 0.4 & $-21.3$ & 10.3 & 26.1 & 170 \\
15:55:43.91 & +11:11:54.4 & 22.9 & 21.6 & 20.8 & g & 0.5963 & 0.5 & $-21.9$ & 10.7 & 32.7 & 218 \\
15:55:45.22 & +11:11:20.7 & 23.3 & 21.9 & 21.1 & g & 0.6626 & 0.4 & $-22.2$ & 10.6 & 32.3 & 226 \\
15:55:45.38 & +11:11:09.9 & 23.6 & 23.1 & 22.9 & s & 1.1130 & 0.2 & $-22.5$ & 10.3 & 37.2 & 305 \\
15:55:42.08 & +11:10:32.1 & 23.1 & 22.2 & 22.0 & g & 0.5244 & 0.2 & $-20.2$ & 9.4 & 54.1 & 338 \\
15:55:44.62 & +11:12:05.5 & 23.7 & 23.2 & 22.8 & g & 0.7753 & 0.2 & $-20.4$ & 9.5 & 47.3 & 351 \\
15:55:40.28 & +11:10:55.3 & 23.9 & 23.1 & 22.3 & g & 0.8778 & 0.4 & $-21.9$ & 10.5 & 49.9 & 386 \\
15:55:47.24 & +11:11:11.6 & 23.7 & 23.1 & 22.3 & g & 0.7270 & 0.2 & $-20.8$ & 9.7 & 63.0 & 457 \\
15:55:37.80 & +11:11:43.4 & 23.0 & 21.4 & 20.9 & g & 0.4695 & 0.4 & $-21.2$ & 10.3 & 79.5 & 469 \\
15:55:38.87 & +11:11:54.8 & 22.5 & 21.6 & 21.1 & g & 0.6935 & 0.2 & $-21.9$ & 10.1 & 68.5 & 488 \\
15:55:37.26 & +11:11:23.9 & 23.2 & 21.5 & 20.9 & g & 0.4699 & 0.5 & $-21.3$ & 10.5 & 85.2 & 503 \\
\vdots      & \vdots        & \vdots   &    \vdots &    \vdots &    \vdots  &  \vdots   &   \vdots &    \vdots      & \vdots  & \vdots \\

\hline
\multicolumn{11}{c}{$z_{\rm gal}=0.4291-0.4387$} \\
\hline
15:55:43.51 & +11:11:13.0 & 23.9 & 22.7 & 22.5 & g & 0.4300 & 0.5 & $-20.8$ & 10.2 & 13.2 & 74 \\
15:55:43.19 & +11:11:02.8 & 23.6 & 22.2 & 21.5 & g & 0.4347 & 0.6 & $-20.5$ & 10.3 & 21.6 & 122 \\
15:55:41.45 & +11:11:29.1 & 25.4 & 23.3 & 22.9 & s & 0.4297 & 0.4 & $-19.1$ & 9.5 & 23.9 & 134 \\
15:55:42.76 & +11:11:56.4 & 24.9 & 23.1 & 22.6 & g & 0.4344 & 0.7 & $-19.6$ & 9.9 & 32.3 & 183 \\
15:55:43.95 & +11:11:56.7 & 21.4 & 19.6 & 18.9 & g & 0.4343 & 0.8 & $-23.2$ & 11.4 & 35.0 & 198 \\
15:55:42.94 & +11:10:47.6 & 23.6 & 22.1 & 21.5 & g & 0.4329 & 0.6 & $-20.4$ & 10.2 & 36.8 & 207 \\
15:55:39.87 & +11:11:46.8 & 24.3 & 23.3 & 23.4 & g & 0.4347 & 0.0 & $-18.5$ & 8.7 & 51.9 & 293 \\
15:55:46.15 & +11:10:54.1 & 22.5 & 21.1 & 20.6 & g & 0.4328 & 0.5 & $-21.3$ & 10.5 & 54.8 & 308 \\
15:55:46.69 & +11:11:07.7 & 23.1 & 22.3 & 22.2 & g & 0.4332 & 0.0 & $-19.5$ & 9.1 & 56.1 & 316 \\
15:55:47.25 & +11:11:16.0 & 22.3 & 20.6 & 20.0 & g & 0.4328 & 0.7 & $-22.1$ & 11.0 & 62.4 & 351 \\
15:55:42.66 & +11:12:29.8 & 23.4 & 22.1 & 21.9 & g & 0.4335 & 0.3 & $-20.0$ & 9.6 & 65.7 & 370 \\
15:55:42.53 & +11:12:30.2 & 24.5 & 22.8 & 22.4 & g & 0.4339 & 0.5 & $-19.7$ & 9.8 & 66.3 & 374 \\
15:55:37.79 & +11:10:51.4 & 24.3 & 23.0 & 22.8 & g & 0.4344 & 0.3 & $-19.2$ & 9.3 & 84.0 & 474 \\
15:55:44.56 & +11:09:57.9 & 24.4 & 23.0 & 22.7 & g & 0.4327 & 0.4 & $-19.5$ & 9.5 & 89.3 & 503 \\
\vdots      & \vdots        & \vdots   &    \vdots &    \vdots &    \vdots  &  \vdots   &   \vdots &    \vdots      & \vdots  & \vdots & \vdots \\
\hline
\multicolumn{11}{c}{$z_{\rm gal}=0.3506-0.3596$} \\
\hline
15:55:44.49 & +11:09:19.3 & 22.8 & 21.7 & 21.4 & g & 0.3531 & 0.4 & $-19.9$ & 9.7 & 126.8 & 630 \\
15:56:02.12 & +11:12:44.1 & 22.7 & 21.6 & 21.3 & g & 0.3537 & 0.4 & $-20.0$ & 9.8 & 291.8 & 1451 \\
15:55:29.10 & +11:05:37.7 & 24.3 & 23.3 & 23.2 & g & 0.3547 & 0.1 & $-18.3$ & 8.6 & 402.8 & 2007 \\
15:55:47.81 & +11:04:41.8 & 22.8 & 21.9 & 21.6 & g & 0.3596 & 0.4 & $-19.8$ & 9.6 & 408.6 & 2054 \\
15:55:41.14 & +11:03:51.8 & 21.2 & 19.5 & 18.8 & g & 0.3580 & 0.8 & $-22.7$ & 11.2 & 453.4 & 2273 \\
15:56:13.90 & +11:09:06.0 & 24.1 & 23.3 & 23.4 & g & 0.3590 & 0.0 & $-18.1$ & 8.5 & 474.8 & 2384 \\
15:55:24.61 & +11:18:04.5 & 22.2 & 20.7 & 20.3 & g & 0.3538 & 0.6 & $-21.1$ & 10.4 & 483.4 & 2405 \\
15:55:12.93 & +11:17:48.8 & 21.9 & 20.9 & 20.6 & g & 0.3542 & 0.4 & $-20.7$ & 10.0 & 586.6 & 2920 \\
15:55:52.15 & +11:21:42.4 & 20.6 & 19.4 & 18.9 & g & 0.3531 & 0.6 & $-22.5$ & 11.0 & 632.4 & 3142 \\
\vdots      & \vdots        & \vdots   &    \vdots &    \vdots &    \vdots  &  \vdots   &   \vdots &    \vdots      & \vdots  & \vdots & \vdots \\
\hline
\multicolumn{11}{c}{Other redshifts} \\
\hline
15:55:07.77 & +11:01:42.3 & 21.1 & 19.8 & 19.3 & g & 0.0017 & 1.2 & $-9.6$ & 5.8 & 780.0 & 28 \\
15:55:46.11 & +11:11:49.4 & 23.5 & 22.9 & 22.8 & g & 0.1022 & 0.3 & $-15.6$ & 7.8 & 51.6 & 97 \\
15:55:44.01 & +11:11:09.1 & 23.1 & 22.6 & 22.3 & g & 0.3892 & 0.3 & $-19.2$ & 9.3 & 20.8 & 110 \\

\vdots      & \vdots        & \vdots   &    \vdots &    \vdots &    \vdots  &  \vdots   &   \vdots &    \vdots      & \vdots  & \vdots & \vdots \\

\hline

\end{tabular}
\end{table*}

\section{Discovery and redshift of the group hosting \es} \label{section:redshift}

{\color{black} Optical$-$X-ray spectra of \es\ exhibit no detected emission lines}, preventing systemic redshift measurements \citep[][]{Landoni:2014}. The lack of a precise redshift measurement complicates the interpretation of absorption features in the spectrum of \es\ due to an inability to differentiate intervening IGM/CGM systems from {\color{black} associated absorption}. 
Previous estimates of the redshift of \es\  based on the detection of intervening H\,I \lya\  absorption \citep[e.g.][]{Danforth:2010} and the shape of its $\gamma$-ray spectrum \citep[e.g.][]{Abramowski:2015} imply $0.413 \lesssim z_{\rm sys} \lesssim 0.6$. {\color{black}

Blazars are typically hosted by luminous elliptical galaxies \citep[e.g.][]{Urry:2000} in massive groups \citep[e.g.][]{Wurtz:1997}. Moreover, \es\ is a high energy peaked blazar which are thought to arise from beamed FR-I radio galaxies \citep[e.g.][]{Rector:2000}. \es\ exhibits a complex, one-sided radio-jet morphology \citep[see Figure \ref{figure:field_image};][]{Rector:2003}, indicating disturbance by a hot intragroup or intracluster medium.} {\color{black} Identification} of the blazar's host group therefore represents a precise means of inferring its redshift \citep[e.g.][]{Rovero:2016, Farina:2016}.

\begin{figure*}
\centering
\includegraphics[width=\textwidth]{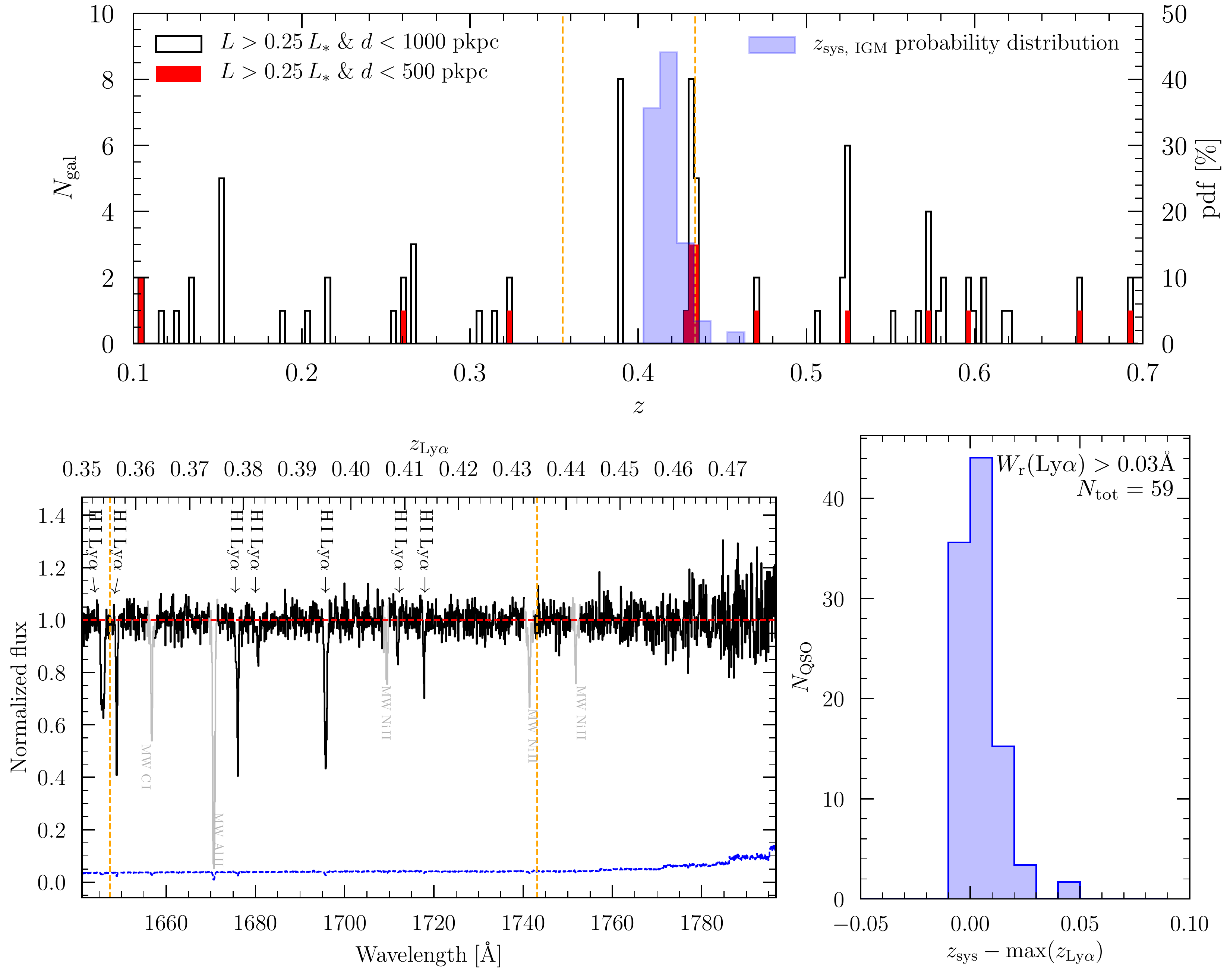}
\caption{{\it Top}: Redshift histograms of galaxies with $L>0.25 L_*$  at $d<1000$ (black histogram) and $<500$ (red filled histogram) pkpc from the \es\ sightline. The only massive group in the field with multiple galaxies of $L>0.25 L_*$ near the blazar sightline is at $z=0.433$, a strong indication that \es\ is a member of this galaxy group. {\it Bottom left}: Continuum normalized COS red-end spectrum of \es\ with flux in black and error in blue. Intervening H\,I \lya\ absorption systems from the IGM are labeled, {\color{black} and Milky Way features are plotted in grey}. The bottom axis shows the observed-frame wavelength while the top axis shows the corresponding \lya\ redshift. Orange dashed lines mark the redshifts of the candidate O\,VII systems. 
{\it Bottom right}: Histogram of the redshift difference between QSO systemic redshifts and the highest redshift H\,I \lya\ absorber of $W_{\rm r} > 0.03$ \AA\ cataloged in COS spectra by \cite{Danforth:2016}, $z_{\rm sys} - {\rm max}(z_{\rm Ly\alpha})$. {\color{black} The resulting empirical constraint on the redshift of \es\ is shown in blue in the top panel (right axis).}}
\label{figure:zhist}
\end{figure*}

To identify the host group of \es, the top panel of Figure \ref{figure:zhist} displays the redshift histogram for galaxies of $L>0.25 L_*$ from our survey at $d<500$ and  $<1000$ proper kpc (pkpc) from the blazar sightline. {\color{black} With high completeness levels of $100\%$, ${>}90\%$, and ${>}80$\% for galaxies of $L>1.0, 0.5,$ and $0.25$ $L_*$ respectively at $d<500$ pkpc and $z<0.6$, our redshift survey is sensitive to galaxy groups over the full range of possible systemic redshifts for \es. The only significant overdensity with multiple luminous galaxies near the blazar sightline is at $z\approx0.433$, a strong indication that the blazar is a member of the $z=0.433$ group.}

{\color{black} The blazar host group consists of $7$ ($14$) members of $L>0.25 L_*$ at $d<500$ (1000) pkpc from the blazar and exhibits a light-weighted redshift of $z_{\rm group}=0.433$. Not including the blazar host, the total stellar mass (luminosity) of the group is $\approx 8\times10^{11} M_\odot$ ($18\ L_*$) with $\approx 60\%$ ($50\%$) coming from three massive, quiescent galaxies of $\log M_*/M_\odot > 11.0$.}
The measured line-of-sight velocity dispersion of the group is $\sigma_{\rm group} \approx 300$ \kms, which corresponds to an estimated dynamical mass of $M_{\rm dyn}\sim 2{-}5\times10^{13}\ M_\odot$. {\color{black} Such a massive group is consistent with expectations for the environment of blazars like \es. Assuming that the luminosity of the blazar host galaxy is $L=6 L_*$ \citep[][]{Urry:2000}, the group light-weighted center is $13''$ E.\ (70 pkpc) and $7''$ N.\ (40 pkpc) of the blazar position.}

The presence/absence of H\,I \lya\ absorption in the blazar spectrum as a function of redshift can be used for an independent estimate of the blazar redshift.
The archival COS spectrum of \es\ enables searches for H\,I \lya\ absorption at $\lambda <1796.7$ \AA\ which corresponds to a maximum redshift of $z_{\rm Ly\alpha}=0.478$. In this wavelength range, {\color{black} the $S/N$ is sufficient} to detect absorbers of $W_{\rm r} > 0.03$ \AA\ at $3\sigma$ significance. The spectrum reveals $7$ \lya\ absorbers at $z=0.350-0.413$ implying $z_{\rm sys}\gtrsim 0.413$ but none over the similar interval of $z=0.413-0.478$ (bottom left panel of Figure \ref{figure:zhist}).

To quantify the redshift constraint on \es\ from the \lya\ forest with objects of similar luminosity, we identified $59$ available QSOs with measured systemic redshifts, archival COS spectra, and IGM absorption line identifications from \cite{Danforth:2016}. For each QSO, we computed the difference between the systemic redshift and that of the highest redshift H\,I Ly$\alpha$ line with $W_r>0.03$ \AA\ in the spectrum, $\Delta z = z_{\rm sys} - {\rm max}(z_{\rm Ly\alpha})$. The resulting $\Delta z$ distribution is shown in the bottom right panel of Figure \ref{figure:zhist}. When combined with the highest redshift \lya\ line in the spectrum of \es\ at $z=0.413$ ({\color{black} 50 pMpc from $z=0.433$ where the UV background dominates}), this distribution implies a 95\% confidence interval for the redshift of \es\ of $z_{\rm sys}=0.411-0.435$, consistent with membership of the $z=0.433$ galaxy group. {\color{black} This constraint assumes that blazars and QSOs reside in similar intergalactic environments and is subject to small number statistics in the wings of the distribution. It will be further tested with new {\it HST} NUV spectra (PI: Muzahid, PID: 15835) for improved \lya\ searches}.

\section{Implications for the WHIM} \label{section:environment}

\begin{figure}
\includegraphics[width=\columnwidth]{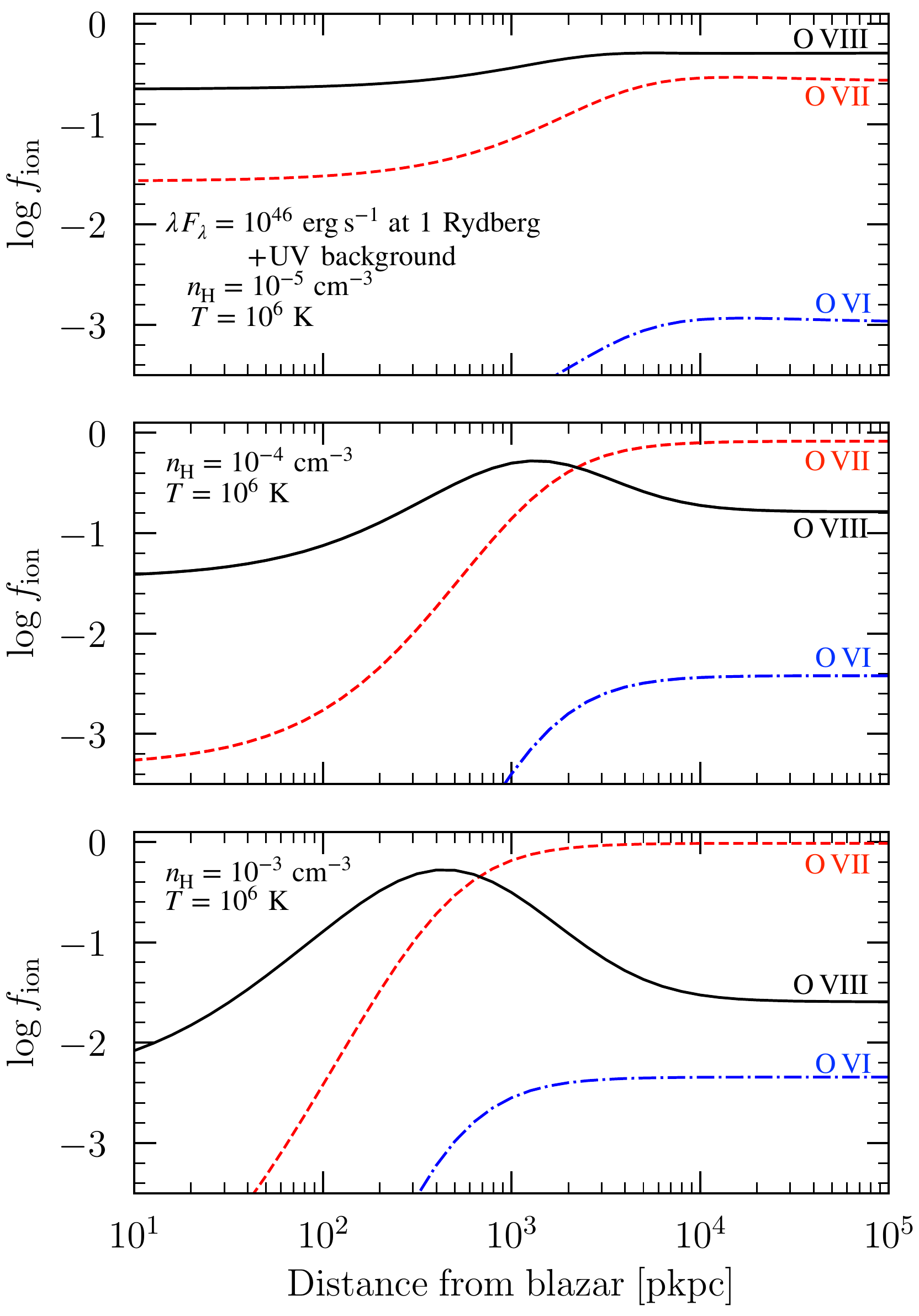}
\caption{Metallicity independent equilibrium ion fraction of O\,VI (blue), O\,VII (red), and O\,VIII (black) as a function of distance from the blazar for gas with a temperature of $T = 10^6$ K and with a density of $n_{\rm H}=10^{-5}$ (top),  $10^{-4}$ (middle), and $10^{-3}$ (bottom) $\rm cm^{-3}$.}
\label{figure:PI_CI}
\end{figure}

\subsection{The origins of candidate WHIM X-ray absorbers} 

{\color{black} \cite{Nicastro:2018} identified two candidate WHIM O\,VII absorbers at $z=0.4339$ and $0.3551$ in the {\it XMM} spectrum of \es}, suggesting that the hot phase of the IGM is metal-rich and potentially closing the missing baryon problem. Neither O\,VII candidate is detected in O\,VI, O\,VIII, or lower ionization metal ions, similar to recent non-detections in an X-ray emitting cosmic filament \citep[][]{Connor:2019}. Here, we discuss the origins of these WHIM candidates based on our redshift survey.

As discussed in Section \ref{section:redshift}, \es\ is most likely a member of a galaxy group at $z=0.433$. The identified O\,VII system at $z=0.4339$ is therefore associated with the blazar environment and cannot be used in cosmic baryon {\color{black} censuses}.


The blazar host group is part of a larger scale overdensity consisting of three additional groups at $\approx 1.5$ pMpc S.E., $\approx1.5$ pMpc N.W., and $\approx 2.5$ pMpc E.S.E. from the blazar. {\color{black} The} O\,VII candidate could be due to WHIM from this overdensity, but photoionization from the blazar and UV background would be important. To evaluate the feasibility of the WHIM interpretation under these circumstances, we ran a series of Cloudy \citep[][]{Ferland:2017} models to calculate the equilibrium O\,VI, O\,VII, and O\,VIII ion fractions as a function of distance from the blazar for gas with $n_{\rm H}=10^{-5}{-}10^{-3}\ {\rm cm}^{-3}$
and temperatures of $T = 10^5-10^7$ K including photoionization from the blazar ($\lambda L_{\lambda}=10^{46}\ {\rm erg\,s^{-1}}$ at $1$ Rydberg and  UV spectral slope of $\alpha = -1.4$ based the COS spectrum) and UV background \citep[][]{Khaire:2019} as shown in Figure \ref{figure:PI_CI}.

\cite{Nicastro:2018} demonstrated that the O\,VII detection and O\,VI/O\,VIII non-detections at $z=0.4339$ require a gas temperature of $T \approx 10^6$ K with little contribution from photoionization. This rules out WHIM gas with $n_{\rm H}<10^{-4}\ {\rm cm^{-3}}$ at any distance from the blazar because photoionization by the UV background is significant at such low densities \citep[see Figure \ref{figure:PI_CI};][]{Wijers:2019}. Denser hot gas of $n_{\rm H}=10^{-4}$ ($10^{-3}$) $\rm cm^{-3}$ can reproduce the absorber properties but only at $>10$ ($1$) pMpc from the blazar (see Figure \ref{figure:PI_CI}). The $z=0.4339$ O\,VII candidate is, therefore, unlikely to be due to low-density WHIM but may arise from hot CGM or intragroup medium \citep[][]{Mulchaey:1996} in the {\color{black} blazar environment}. 
 
The $z=0.3551$ {\color{black} O\,VII candidate} resides in a comparatively isolated region with no galaxies at $d<500$ pkpc from the sightline within $\Delta v = \pm 1000$ \kms\ from the absorber redshift despite $100\%$ completeness levels for galaxies of $L>0.1 L_*$. The nearest galaxy to the sightline is a star-forming galaxy of $\log M_*/M_\odot = 9.7$ at $z=0.3531$ and $d=630$ pkpc or $\approx 5\times$ {\color{black} its virial radius} (estimated with the stellar-to-halo mass relation from \cite{Kravtsov:2018} and virial radius definition from \cite{Bryan:1998}). The nearest massive galaxy has a stellar mass of $\log M_*/M_\odot = 11.2$ and is at $d=2273$ pkpc or $\approx 5\times$ {\color{black} its} virial radius.
The $z=0.3551$ candidate could be due to an undetected dwarf in principle, but surveys of the CGM/IGM around dwarfs \citep[][]{Johnson:2017} indicate that metal absorption systems are rare beyond the virial radius, and dwarfs are not expected to maintain a hot halo \citep[e.g.][]{Correa:2018}. 

To determine whether strong O\,VII systems are expected from the WHIM in isolated environments, we calculated the fraction of predicted strong O\,VII absorbers as a function of environment using WHIM predictions \citep[][]{Wijers:2019} from the EAGLE cosmological hydrodynamical simulations \citep[][]{Schaye:2015, Crain:2015, McAlpine:2016}. We calculated column densities within $2000\ {\rm km\,s}^{-1}$ simulation slices and cross-correlated with galaxies as a function stellar mass and projected distance. In total, only $1-3$\% of the predicted, comparably strong O\,VII ($\log N({\rm O\,VII})/{\rm cm^{-2}}=15.6$) systems occur in similarly isolated environments ($d>630$ pkpc to the nearest galaxy of $\log M_*/M_\odot > 9.7$). While the model predictions are subject to non-negligible uncertainties {\color{black} due to} treatment of peculiar velocities and feedback, we nevertheless conclude that strong O\,VII WHIM systems are not expected to be common in isolated environments. Moreover, \cite{Bonamente:2018} estimated a 4\% probability that the $z=0.3551$ O\,VII candidate arises from noise fluctuations.

{\color{black} We conclude} that neither of the two candidate O\,VII absorbers in the spectrum of \es\ are of confident and unbiased intergalactic origin. This implies a 95\% upper limit on the number of WHIM O\,VII absorbers with $W_{\rm r} \gtrsim 6$ m\AA\ per unit redshift of $\frac{dN}{dz}<8$. The lack of strong WHIM X-ray absorption systems suggests that metal enrichment is primarily confined to galaxy halos and their immediate outskirts. This is consistent with the EAGLE simulations which predict that most strong {\color{black} O\,VII systems arise} from metal rich ($>0.5 Z_\odot$) gas at over-densities of $\delta \gtrsim 100$ \citep[see][]{Wijers:2019}. Further exploration of the relationship between the WHIM and galaxies {\color{black} requires} metallicity independent probes.

\subsection{The origins of broad H\,I \lya\ systems}

While metallicity independent probes of the hot IGM are not currently available (except via stacking), broad H\,I absorbers ($b>40$ \kms) can be used to trace metal poor, warm IGM. {\color{black}While temperature measurements are not possible for most broad H\,I systems due to lack of detected metals, \cite{Savage:2014} found that $78^{+7}_{-12}\%$ of broad H\,I absorbers with well aligned O\,VI detections exhibit warm-hot temperatures of $\log T/{\rm K}=5-6$.} \cite{Danforth:2010} identified 12 broad H\,I {\color{black} absorbers} in the COS spectrum of \es\ at $\Delta v <-10,000$ \kms\ from the blazar redshift. {\color{black} None} of the broad H\,I absorbers are coincident with detected galaxies at $d<R_{\rm h}$. However, {\color{black} all} are coincident with at least one luminous galaxy of $L>0.25 L_*$ within $\Delta v = \pm 1000$ \kms\ with a median projected distance to the closest one of $700$ pkpc. In contrast, narrow ($b<30$ \kms; $T<5\times10^4$ K) H\,I absorption systems detected toward \es\ are further from luminous galaxies on average with a median distance to the nearest one of $1300$ pkpc while O\,VI absorbers are closer to luminous galaxies \citep[$350$ pkpc;][]{Johnson:2013, Pratt:2018}.

\subsection{Summary and conclusions}

Based on deep and highly complete redshift surveys in the field of \es\, we found that: \begin{enumerate}
	\item Neither of the two candidate O\,VII WHIM systems reported toward the sightline \citep[][]{Nicastro:2018} are of confident and unbiased intergalactic origin. {\color{black} The origins, state, and cosmological mass density of the hot IGM therefore remain uncertain}. 
	\item Low metallicity warm IGM traced by broad H\,I \lya\ absorbers occur $\approx 2\times$ further from luminous ($L>0.25 L_*$) galaxies than O\,VI absorbers on average, but $2\times$ closer than {\color{black} cool IGM traced by narrow \lya}.
\end{enumerate}
Our findings are consistent with gravitational collapse heating portions of the IGM to form the WHIM. However, they also suggest that feedback is ineffective at enriching the low-$z$ IGM far beyond galaxy/group halos to levels currently observable in UV and X-ray metal ions. Indeed, \cite{Liang:2014} and \cite{Johnson:2015a} placed upper limits on the mean metallicity of the IGM of $<0.1 Z_\odot$ and pristine ($Z<0.01 Z_\odot$) gas can be found even around massive galaxies \citep[][]{Chen:2019}. These observations highlight the need for {\color{black} a variety of  WHIM probes coupled with deep galaxy surveys}.

\section*{Acknowledgements}

We are grateful to J. Nevalainen, F. Nicastro, and M. Petropoulou for insightful comments.
SDJ is supported by a NASA Hubble Fellowship (HST-HF2-51375.001-A).
MRD acknowledges support from the Dunlap Institute at the University of Toronto and the Canadian Institute for Advanced Research (CIFAR).
J.C.C. acknowledges support by the National Science Foundation under Grant No. AST-1517816.
Based on observations from the Magellan, the NOAO Mayall, and NASA/ESA {\it Hubble} Telescopes. The authors are honored to conduct research on Iolkam Du\'ag (Kitt Peak), a mountain with particular significance to the Tohono O\'odham. 
We made use of the NASA Astrophysics Data System.

\facilities{Magellan, {\it HST}, Mayall}


\end{document}